\documentclass[aps,prl,twocolumn,superscriptaddress,showpacs,showkeys]{revtex4}
\usepackage{graphicx}
\usepackage{gensymb}
\usepackage{color}
\usepackage{ulem}
\usepackage{amsmath,amssymb}
\usepackage{endnotes}

\makeatletter

\def\enoteheading{\section*{\notesname
  \@mkboth{\MakeUppercase{\notesname}}{\MakeUppercase{\notesname}}}%
  \mbox{}\par\vskip-2.3\baselineskip\noindent\rule{.5\textwidth}{0.4pt}\par\vskip\baselineskip}
\makeatother

\newcommand\foreign[1]{\textit{#1}}

\begin{document}
\title{Electronic Properties of Single-Layer CoO$_2$/Au(111)}

\author{Ann Julie U. Holt}
\affiliation{Department of Physics and Astronomy, Interdisciplinary Nanoscience Center (iNANO), Aarhus University, 8000 Aarhus C, Denmark}
\author{Sahar Pakdel}
\affiliation{Department of Physics and Astronomy, Interdisciplinary Nanoscience Center (iNANO), Aarhus University, 8000 Aarhus C, Denmark}
\author{Jonathan Rodr{\'i}guez-Fern{\'a}ndez}
\affiliation{Department of Physics, University of Oviedo, Oviedo 33007 Spain}
\author{Yu Zhang}
\affiliation{Artemis Program, U.K. Central Laser Facility, STFC Rutherford Appleton Laboratory, Didcot, Oxfordshire, OX11 0QX, U.K.}
\author{Davide Curcio}
\affiliation{Department of Physics and Astronomy, Interdisciplinary Nanoscience Center (iNANO), Aarhus University, 8000 Aarhus C, Denmark}
\author{Zhaozong Sun}
\affiliation{Interdisciplinary Nanoscience Center (iNANO), Aarhus University, 8000 Aarhus C, Denmark}
\author{Paolo Lacovig}
\affiliation{Elettra—Sincrotrone Trieste S.C.p.A., AREA Science Park, Strada Statale 14, km 163.5, 34149, Trieste, Italy}
\author{Yong-Xin Yao}
\affiliation{Ames Laboratory U.S.-DOE, Ames, Iowa 50011, U.S.A.}
\affiliation{Department of Physics and Astronomy, Iowa State University, Ames, Iowa 50011, U.S.A.}
\author{Jeppe V. Lauritsen}
\affiliation{Interdisciplinary Nanoscience Center (iNANO), Aarhus University, 8000 Aarhus C, Denmark}
\author{Silvano Lizzit}
\affiliation{Elettra—Sincrotrone Trieste S.C.p.A., AREA Science Park, Strada Statale 14, km 163.5, 34149, Trieste, Italy}
\author{Nicola Lanat\`a}
\affiliation{Department of Physics and Astronomy, Interdisciplinary Nanoscience Center (iNANO), Aarhus University, 8000 Aarhus C, Denmark}
\affiliation{Nordita, KTH Royal Institute of Technology and Stockholm University, Roslagstullsbacken 23, 10691 Stockholm, Sweden}
\author{Philip Hofmann}
\affiliation{Department of Physics and Astronomy, Interdisciplinary Nanoscience Center (iNANO), Aarhus University, 8000 Aarhus C, Denmark}
\author{Marco Bianchi}
\affiliation{Department of Physics and Astronomy, Interdisciplinary Nanoscience Center (iNANO), Aarhus University, 8000 Aarhus C, Denmark}
\author{Charlotte E. Sanders}
\email{charlotte.sanders@stfc.ac.uk}
\affiliation{Artemis Program, U.K. Central Laser Facility, STFC Rutherford Appleton Laboratory, Didcot, Oxfordshire, OX11 0QX, U.K.}

\date{\today}
\begin{abstract}

We report direct measurements via angle-resolved photoemission spectroscopy (ARPES) of the electronic dispersion of single-layer CoO$_2$.  The Fermi contour consists of a large hole pocket centered at the $\overline{\Gamma}$ point.  To interpret the ARPES results, we use density functional theory (DFT) in combination with the multi-orbital Gutzwiller Approximation (DFT+GA), basing our calculations on crystalline structure parameters derived from x-ray photoelectron diffraction and low-energy electron diffraction. Our calculations are in good agreement with the measured dispersion.  We conclude that the material is a moderately correlated metal.  We also discuss substrate effects, and the influence of hydroxylation on the CoO$_2$ single-layer electronic structure.

\end{abstract}


\maketitle

\section{I. INTRODUCTION}

Layered bulk crystals based on hexagonal CoO$_2$ \cite{Takada:2003, Sakurai:2015, Raveau:2015} display intriguing electronic and phononic properties that arise from the quasi-two-dimensional (quasi-2D) nature of the atomic layers.  For example, when the CoO$_2$ layers are interleaved with H$_2$O, Na$^+$, and H$_3$O$^+$, superconductivity is observed at transition temperatures T$_C$ of approximately 4--5\,K\,\cite{Takada:2003,Sakurai:2015}, and the nearly 2D character of the CoO$_2$ layers is understood to be a key aspect of these superconducting properties \cite{Takada:2003, Lorenz:2003, Milne:2004, Wang:2005}. Similarities to the high-$T_C$ copper oxides are also notable.  In both cases, a strong anisotropy between the in-plane and out-of-plane directions is key to the materials' electronic properties \cite{Nakamura:1993, Terasaki:1997, Basov:1999, Sugiura:2009}.  Atomic layers that intervene between oxide layers play, in both cases, a complex role that goes beyond simple doping to determine the special many-body physics of the whole system \cite{Yang:2005, Geballe:2012, Takahata:2000} (although the superconducting phases certainly also have a critical dependence on doping \cite{Schaak:2003, Foo:2004, Rybicki:2016}).

Given that the complex electronic properties of these bulk systems arise as quasi-2D physics in weakly interacting, atomically thin oxide layers, it is natural to ask whether any of the interesting electronic properties of the bulk persist in the single-layer (SL) limit.  The electronic properties of a SL material can differ in important ways from those of layered bulk parent compounds.  For example, among the transition metal dichalcogenides, the band dispersion \cite{Kuc:2011,Zhu:2011} and electronic correlations \cite{Feng:2018,Xi:2015} can be significantly modified in the SL limit.

A method has recently been developed for epitaxially fabricating rotationally aligned SL CoO$_2$ islands on Au(111), Pt(111), and Ag(111) substrates \cite{Walton:2015,Fester:2017a}. So far, the SL system has been studied in the context of applications to catalysis \cite{Fester:2018b}.  Here we investigate the electronic properties of the SL, which---to the best of our knowledge---have not yet been studied experimentally, although a recent theory paper has predicted that the SL might manifest 2D ferromagnetism and undergo a superconducting transition at $T_C$\,=\,25--28\,K\,\cite{Nguyen:2019}.  Besides laying the groundwork for new directions in the very active field of 2D-materials research, our study aims to clarify our understanding of layered CoO$_2$-based compounds.

\section{II. METHODS}
 
Our growth procedure is based on the synthesis method that has been previously described by Walton, \textit{et al.} \cite{Walton:2015}. It consists of two steps, performed \textit{in-situ} in a vacuum chamber with a base pressure of low-$10^{-10}$\,mbar. First, we evaporate elemental Co while simultaneously exposing the sample to O$_2$ at a chamber pressure of $6\cdot 10^{-7}$\,mbar.  In this step, the sample temperature is ca. 380\,K. This forms SL CoO on the Au(111) substrate. We use a growth rate of approximately one monolayer CoO per hour. We then stop depositing Co and increase the local O$_2$ pressure, using a moveable O$_2$ doser that we bring to within a few mm of the sample face:  with this we further oxidize the CoO for two hours, to form CoO$_2$. The sample temperature is in this second step is ca. 325\,K and the chamber pressure is $4\cdot 10^{-6}$\,mbar, but the local pressure at the sample face is presumably much higher (likely as much as two orders of magnitude). The full procedure results in SL CoO$_2$ islands \cite{Walton:2015}.

Electronic structure measurements were made by angle-resolved photoemission spectroscopy (ARPES) at the SGM3 beamline of the ASTRID2 Synchrotron Light Source in Denmark \cite{Hoffmann:2004}.  
Incident light is linearly polarized in the direction parallel to the scattering plane; the angle between analyzer axis and incident light axis is 50\degree.  Sample quality and coverage were assessed \textit{in situ} via a combination of x-ray photoelectron spectroscopy (XPS), low-energy electron diffraction (LEED), and scanning-tunneling microscopy (STM). The sample was at room temperature during STM characterisation \cite{Supplement}, while it was at 30(5)\,K during XPS, LEED and ARPES measurements.  The width of the Fermi edge in the ARPES measurements (fitted from projected bulk continuum states of the Au substrate) was approximately 80\,meV.

In order to interpret the ARPES measurements, a precise knowledge of structural parameters is crucial.  Previous studies of SL CoO$_2$ on noble metal (111) surfaces \cite{Walton:2015, Fester:2017a, Fester:2017b} have used STM and XPS to obtain structural information about the SL.  However, XPS is a rather indirect method of assessing structure.  STM does provide structural information, but mainly only about the top layer of atoms, and it is limited in its scope to local measurements, so it is not an optimal probe of average atomic structure over macroscopic areas. Here we use x-ray photoelectron diffraction (XPD) as a direct, high-resolution probe of the local geometric structure around each chemically distinct type of emitter, averaged over the area of the beam spot. Measurements were made at the SuperESCA beamline at Elettra, the synchrotron radiation facility in Trieste, Italy \cite{Baraldi:2003}. Here the incident light is linearly polarized in the horizontal plane, where also the electron energy analyser lies, at an angle of 70\degree\,with respect to the photon beam. The overall energy resolution was below 100\,meV and 200\,meV in the energy range from 650\,eV to 1150\,eV photon energy, respectively. The high resolution spectra of O\,1s and Co\,2p core levels were measured in normal emission conditions and the binding energy was normalized to the Fermi level of the Au substrate. Sample cleanliness, layer quality and order where checked with XPS and LEED prior to the XPD measurements. 

XPD patterns are constructed by collecting XPS spectra across a range of polar ($\theta$) angles, from 70\degree (normal incidence in the configuration of SuperESCA) to normal emission, and azimuthal angles ($\phi$) across a range of 130\degree, with an angular resolution in the order of 1\degree. More details can be found in \cite{Bana:2018} and \cite{Bignardi:2019}. For the measurements in this study, the sample was at room temperature. 

In the peak fitting analysis of the resulting data, the photoemission intensity $I$($\theta$, $\phi$) of each component is extracted from every XPS spectrum by the use of a Doniach-Sunjic fitting function \cite{Doniach:1970} with Shirley background subtraction. From this we obtain the modulation function $\chi$, defined as
\begin{equation}
    \chi = \frac{I(\theta, \phi) - I_0(\theta)}{I_0(\theta)},
\label{eq:Chi}
\end{equation}
where $I_0(\theta)$ is the average intensity for an azimuthal scan at polar angle $\theta$. The XPD pattern is a projection of the modulation function for a particular peak. We compare measured XPD patterns to multiple scattering simulations for trial structures that are generated using the program package Electron Diffraction in Atomic Clusters (EDAC) \cite{Garcia:2001}. The agreement between measured and simulated XPD patterns is then quantified, via calculation of a reliability factor $R$ that is defined as the sum of the normalised mean-square deviation of the experimental ($\chi_{ex}$) and theoretical ($\chi_{th}$) modulation functions,
\begin{equation}
    R = \frac{\sum_i (\chi_{th,i} - \chi_{ex,i})^2}{\sum_i (\chi_{th,i}^2 + \chi_{ex,i}^2)},
\end{equation}
for each emission angle $i$. By this definition, an R-factor of 0 corresponds to perfect agreement, while an R-factor of 1 indicates uncorrelated data \cite{Woodruff:2007}. (Anti-correlated data give an R-factor of 2.) Values of R less than 0.3  are generally taken to indicate relatively good agreement (\textit{e.g.}, \cite{Bana:2018}). We take the confidence interval $\Delta R$ of the minimized $R$ value to be $\Delta R = R_{min} \sqrt{2/N} \sim 0.01$ \cite{Bignardi:2019,Bana:2018}, with $N \approx 250$ being the ratio of the solid angle of the measurements to the solid-angular resolution. 

\section{III. RESULTS}

\textbf{Electronic structure determination.}

\begin{figure}
\includegraphics[width=0.48\textwidth]{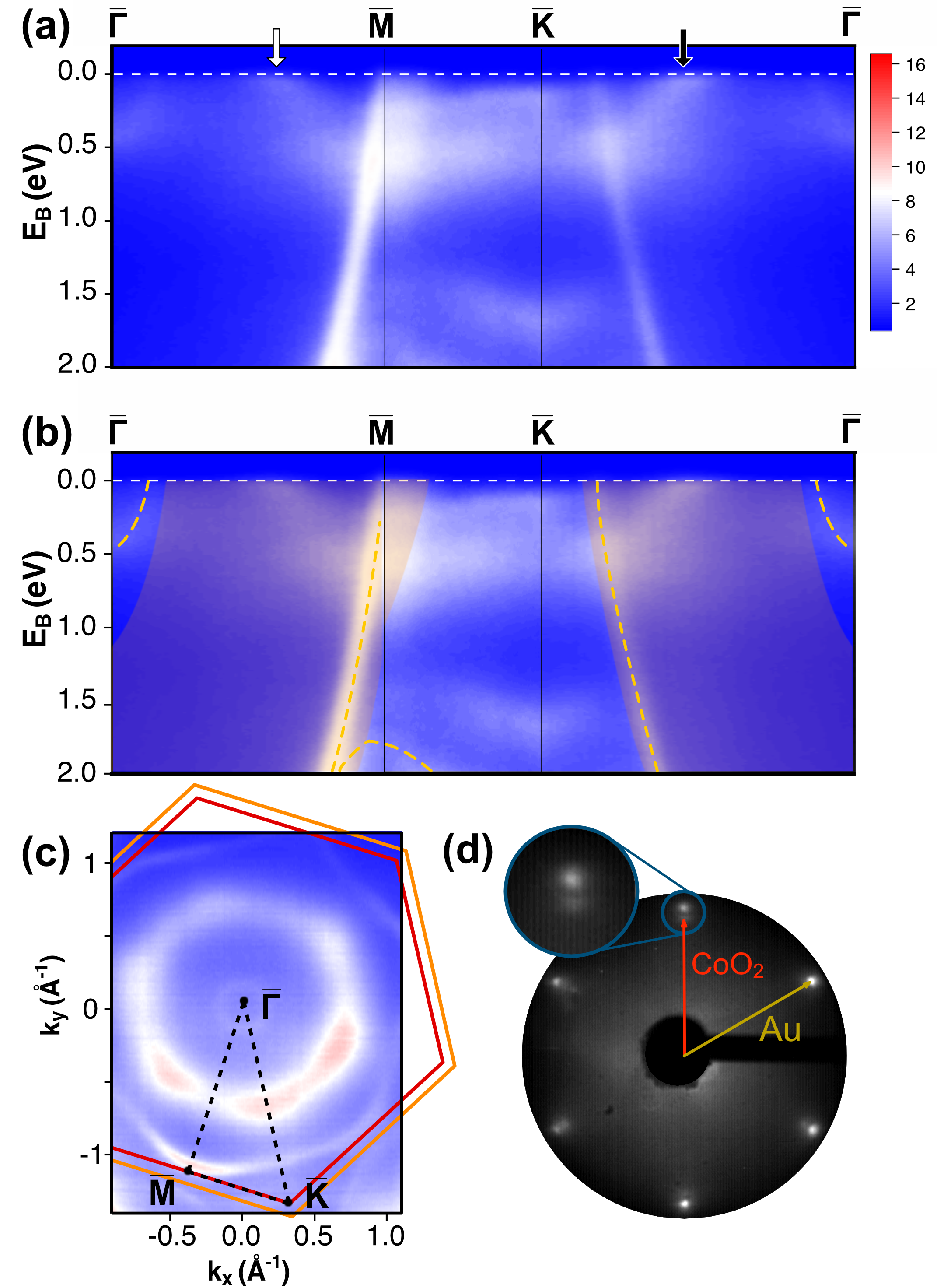}\\
\caption{(a) ARPES data showing the electronic dispersion of 0.7\,SL CoO$_2$ on Au(111), acquired with photon energy $h\nu$\,=\,89\,eV.  High symmetry directions refer to the BZ of CoO$_2$. (b) Same as (a), with relevant features from the Au substrate overlaid in orange as guides to the eye. Dashed lines mark surface state and bulk bands, and the projected bulk band continuum is indicated with shading. (c) Photoemission intensity at the Fermi level. Au(111) and CoO$_2$ BZs are marked by an orange and a red hexagon, respectively, and the high symmetry directions in (a) are marked by black dashed lines. (d) Typical LEED measurement (same sample as shown in Figs. \ref{fig:XPS} and \ref{fig:XPD}). This particular measurement was acquired at SuperESCA using a kinetic energy of 61\,eV, sample temperature approximately 290\,K.}
\label{fig:ARPES}
\end{figure}

ARPES measurements acquired with photon energy $h\nu$\,=\,89\,eV are presented in Fig.~\ref{fig:ARPES} for SL CoO$_2$ with a sample coverage of approximately 70\%, as determined by \foreign{in-situ} STM (see \cite{Supplement}). Fig.~\ref{fig:ARPES}(a) shows the APRES spectrum along cuts through the high-symmetry points $\overline{\Gamma}$, $\overline{\mbox{M}}$, and $\overline{\mbox{K}}$ of the CoO$_2$ Brillouin zone (BZ). The Au(111) surface state at $\overline{\Gamma}$ is faintly visible---presumably deriving from exposed regions of the Au(111) substrate---and so are the sharp, highly dispersive features of the bulk Au(111) \textit{sp}-band, which cross the Fermi level near $\overline{\mbox{M}}$ and along the $\overline{\Gamma}\overline{\mbox{K}}$ cut \cite{Takeuchi:1991}. In Fig.~\ref{fig:ARPES}(b), the electron dispersion from Fig.~\ref{fig:ARPES}(a) is reproduced with the Au features highlighted:  the Au surface state and bulk bands are overlaid with orange dashed lines as guides to the eye, and the projected bulk band continuum is shaded orange. Several observed features are not associated with the Au(111) substrate, and we attribute these to the electronic structure of SL\,CoO$_2$. At first glance, the adlayer band structure shows a metallic character and seems to consist of one band around 1.5\,eV binding energy E$_B$ (mostly visible between $\overline{\mbox{M}}\overline{\mbox{K}}$), and at least two bands close to the Fermi level, at less than 1.0\,eV in binding energy. Of these bands close to the Fermi level, the lower one appears to exhibit energy minima at approximately E$_B$\,=\,0.6\,eV at $\overline{\mbox{M}}$ and approximately E$_B$\,=\,0.5\,eV close to $\overline{\mbox{K}}$.  (As we will see below, however, the actual dispersion is more complicated than this.) A Fermi level crossing at $k_\parallel$\,=\,0.67(3)\,{\AA}$^{-1}$  in the $\overline{\Gamma}\overline{\mbox{M}}$ direction is marked with a white arrow in the figure.  In the $\overline{\Gamma}\overline{\mbox{K}}$ direction, the band crosses the Fermi level at approximately $k_\parallel$\,=\,0.69(5)\,{\AA}$^{-1}$ (black arrow). To identify the Fermi level crossing, the band dispersion near the Fermi level was determined by fitting energy distribution curves (EDCs) to Gaussian-broadened Lorentzians convoluted with the Fermi function, within a range of about 100\,meV below the Fermi level.  The EDC fits were also found to agree well with Gaussian fits to momentum distribution curves (MDCs) within the same energy range. The bands appear to flatten close to the Fermi level, creating the impression of kinks in the dispersion; this will be further discussed below. The Fermi contour corresponding to the measured dispersion is presented in Fig.~\ref{fig:ARPES}(c). The overlaid orange hexagon marks the Au(111) surface BZ (SBZ), which is slightly larger than the CoO$_2$ BZ (red hexagon). The high-symmetry directions in Fig.~\ref{fig:ARPES}(a) are marked with black dashed lines, and the high-symmetry points are labeled. The Au(111) surface state creates a small ring around $\overline{\Gamma}$, and the large, faint hexagonal feature with its vertices at approximately the $\overline{\mbox{M}}$ points of CoO$_2$ arises from Au bulk states.  Between these, the SL CoO$_2$ forms a distorted, rounded hexagon centered around $\overline{\Gamma}$; by referencing the dispersion in panel (a), one can see that this is a large hole pocket. The relative sizes and rotations of the BZs are calculated on the basis of LEED measurements, and a representative measurement is presented in Fig.~\ref{fig:ARPES}(d). Diffraction spots forming two rotationally aligned hexagons arise from the SL CoO$_2$ (red arrow) and the Au(111) substrate (yellow arrow). By using the known in-plane Au(111) lattice parameter of 2.88\,{\AA}~\cite{Maeland:1964}, the in-plane lattice parameter of the SL CoO$_2$ is determined to 3.12(3)\,{\AA}, which is within the range of values (2.8--3.3\,{\AA}) previously reported on the basis of STM measurements~\cite{Walton:2015, Fester:2017b}. \\

\textbf{Core level data.}

\begin{figure*}
\includegraphics[width=\textwidth]{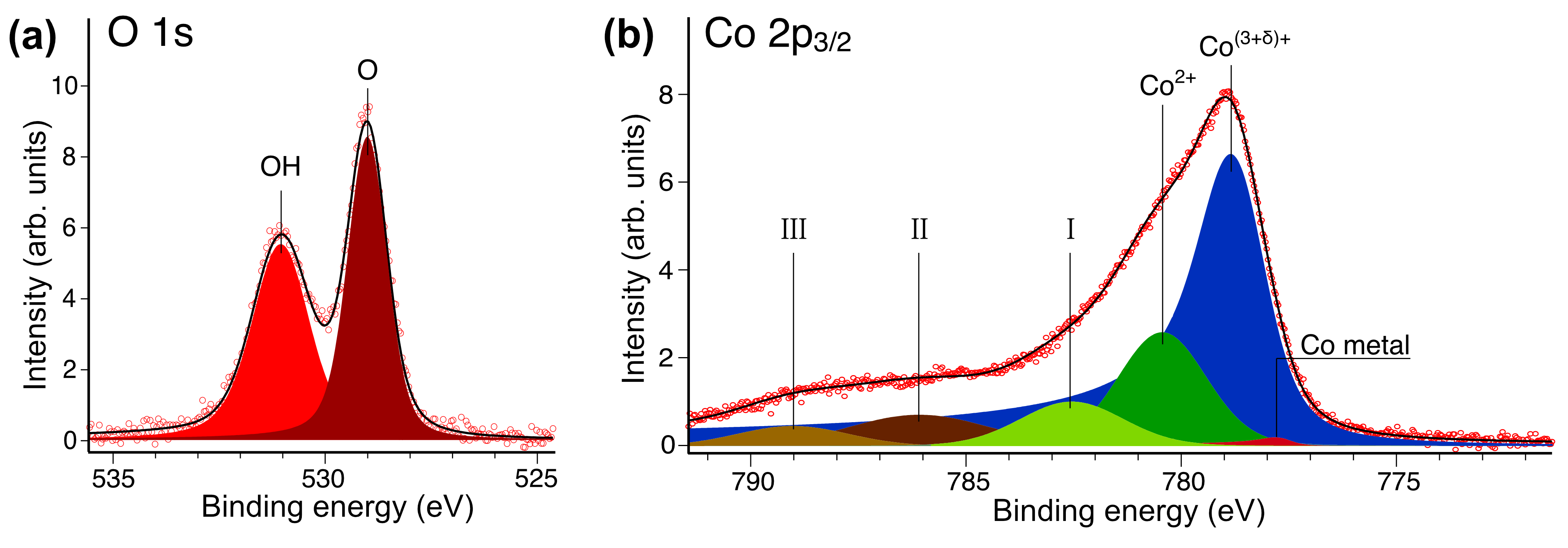}
\caption{XPS spectra from a sample with approximately 55\% SL CoO$_2$ coverage of the Au(111) surface, measured in normal emission conditions, with the Shirley background subtracted. (Coverage was estimated from the quenching of the Au\,4f surface-shifted core-level peak component \cite{Supplement,Heimann:1981}.) (a) O\,1s core level, acquired with photon energy $h\nu$\,=\,650\,eV, and (b) Co\,2p, measured with $h\nu$\,=\,1150\,eV. The deconvoluted components obtained by curve fitting are  shown. A peak arising from the Co$^{2+}$ multiplet splitting as well as two satellite peaks associated with the Co$^{2+}$ and Co$^{(3+\delta)+}$ components are labelled ``I'', ``II'' and ``III,'' respectively. All the peaks shown arise from Co\,2p$_{3/2}$ core levels; spin-orbit splitting places the Co\,2p$_{1/2}$ peak at approximately 15\,eV higher binding energy than its Co\,2p$_{3/2}$ partner, and hence past the left edge of the plot \cite{XPSHandbook}.}
 \label{fig:XPS}
\end{figure*}

The O\,1s spectrum is displayed in Fig.~\ref{fig:XPS}(a). The main component (dark red) occurs at a binding energy of E$_B$\,=\,529.0(1)\,eV (binding energies are referenced to the measured Fermi level). An additional hydroxyl component (bright red) is also present, as expected from previous studies \cite{Fester:2017b, Fester:2018a, Fester:2018b}.  Similarly to what is reported in the earlier-published results, this OH peak is shifted 2.05(5)\,eV towards higher binding energy relative to the main O peak. Fester, \foreign{et al.}, attribute the presence of this component to a partial hydroxyl overlayer formed by H bonding to O at the top of the CoO$_2$ SL (\foreign{i.e.,} the side of the SL away from the interface with the Au(111) substrate).  They suggest that the hydroxylation results from dissociative adsorption of H$_2$O or H$_2$ rest gas in the vacuum chamber \cite{Fester:2017b}. The main O peak contains photoemission intensity from unhydroxylated emitters at both the top and the bottom of the layer. These components are known to be very closely separated in energy \cite{Fester:2018a} and cannot be resolved.

Fig.~\ref{fig:XPS}(b) shows the Co\,2p spectrum. It is dominated by a large peak with several closely spaced minor components. The complexity of this spectrum attests to the presence of more than one oxidation state of Co. Freestanding CoO$_2$ might be na{\"i}vely expected to exhibit the Co$^{4+}$ oxidation state; however, we instead observe a Co$^{(3+\delta)+}$ oxidation state---fitted here with an asymmetric main peak (blue) at a binding energy of E$_B$\,=\,778.7\,eV---consistent with previously published findings \cite{Walton:2015, Fester:2017b}.  The $(3+\delta)^{+}$ oxidation state is presumably due to charge transfer from adsorbates (or possibly from the substrate, but see discussion below).  We surmise that such charge transfer stabilises the SL; however, the possible mechanism for such stabilisation is beyond the scope of the present work.  Some areas of CoO have failed to further oxidize in the second step of the growth procedure, and this is manifest in the persistent presence of a set of peaks arising from the Co$^{2+}$ oxidation state. Co$^{2+}$ has a high-spin configuration that generates a complex peak structure with multiplet splitting and shake-up satellites \cite{Frost:1974,Chuang:1976,Biesinger:2011,Kim:1975}.  Here we fit it with just two components, aside from the Co$^{2+}$ main component at E$_B$\,=\,780.4(1)\,eV (dark green in the figure):  a shake-up satellite, labelled ``II'' (dark brown), separated from the main peak by 5.7(1)\,eV, and peak ``I'' (light green), which arises from multiplet splitting and is fixed to a separation of 2.1\,eV from the main peak \cite{Biesinger:2011}.  While the intensities and locations of these two peaks are in relatively good agreement with related peak structures previously identified in the literature \cite{Biesinger:2011,Kim:1975,Chuang:1976}, our fitting here is phenomenological and does not attempt to capture the details of the complex peak structure of the Co$^{2+}$ spectrum. Note that this makes it difficult for us to estimate the precise amount of the less-oxidized CoO that remains on this particular sample after the second oxidation step; however, it does not impact the modulation of the Co$^{(3+\delta)+}$ component, which is only associated with photoemission from fully oxidized CoO$_2$.

We identify an additional weak satellite peak (``III," shown in light brown) at 10.3(1)\,eV higher binding energy than the Co$^{(3+\delta)+}$ peak. The position and the intensity of this satellite are similar to what has been seen in diverse related compounds such as CoOOH, Co(OH)$_2$, and Co$_3$O$_4$ \cite{Chuang:1976, Yang:2010, Biesinger:2011}, though its physical interpretation remains uncertain.    Finally, a negligible amount of unoxidized Co metal remains, even after the full growth procedure, and leads to the small peak shown in red in the figure\endnote{The Co\,2p$_{3/2}$ spectrum for bulk metallic Co is known to display an asymmetric main peak together with surface and bulk plasmon peaks \cite{Grosvenor:2005}.  Here, the plasmon satellites are excluded from the metallic Co fit, due to the very small amount of metallic Co found to be present. The binding energy is set to match the value reported by Walton, \foreign{et al.} \cite{Walton:2015}, in their measurements of Co metal on Au(111), and the peak shape is asymmetric with both the shape and location similar to what is measured by Walton \textit{et al.}\,\cite{Walton:2015}.}.\\

\textbf{Crystalline structure determination.}

\begin{figure}
\includegraphics[width=0.48\textwidth]{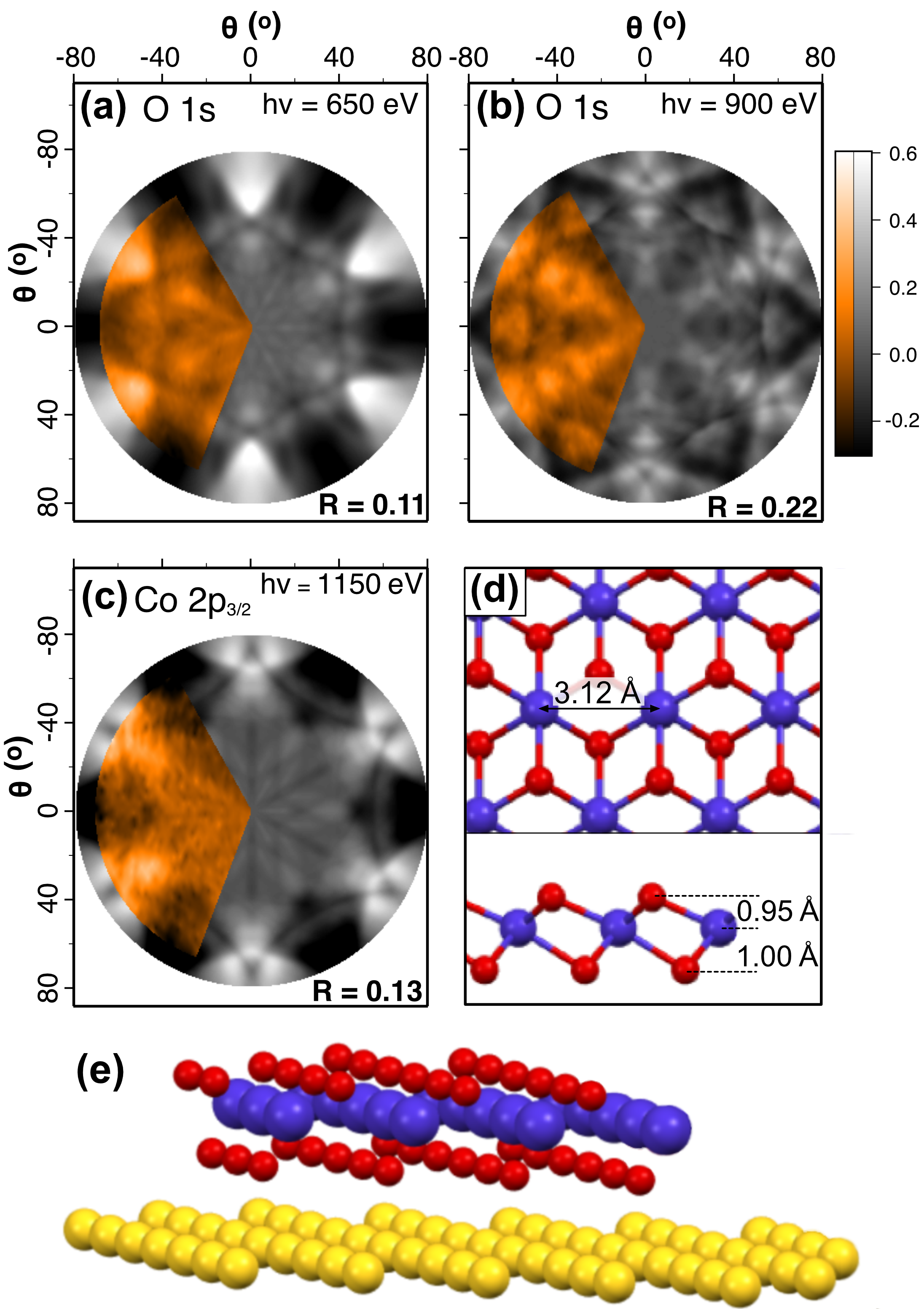}
\caption{XPD patterns from the primary O\,1s component in the (a) backscattering ($h\nu$\,=\,650\,eV) and (b) forward-scattering ($h\nu$\,=\,900\,eV) regimes. (c) XPD from the Co\,2p$_{3/2}$ ${(3+\delta)+}$ component in the forward-scattering regime ($h\nu$\,=\,1150\,eV). Azimuthal manipulator angle corresponds to azimuthal angle in plot; polar manipulator angle $\theta$ corresponds to radial direction in plot. Experimental measurements are shown in orange and are superimposed on their associated best-fit simulations, which are shown in grayscale. (d,e) Structural model derived from the best-fit multiple scattering simulations shown. (d) shows top and side views.  (e) shows an oblique view, with a Au(111) surface at an arbitrary distance from the layer, for illustrative purposes.  The substrate was not actually included in the XPD simulations (see text).} 
 \label{fig:XPD}
\end{figure}

XPD results are presented in Fig.~\ref{fig:XPD}.  The experimental data are shown in orange, superimposed on the grayscale best-fit multiple-scattering simulations. Fig.~\ref{fig:XPD}(a) shows the diffraction pattern of the unhydroxylated O\,1s core level peak obtained with a photon energy of $h\nu$\,=\,650\,eV. This photon energy was chosen so that the photoelectron kinetic energy would be less than 150\,eV, to enhance the cross section for backscattering of photoelectrons from the underlying structure and thus the sensitivity of the measurement to emitters in the top layer of O atoms. By contrast, in Fig.~\ref{fig:XPD}(b) a higher photon energy of $h\nu$\,=\,900\,eV was used, favouring the forward scattering of photoelectrons and enhancing sensitivity to emitters in the bottom layer of O atoms. The resulting modulation function is, in this latter case, highly dependent on the structure above the bottom layer of O---the arrangement of Co atoms and of O atoms at the top of the layer. The diffraction pattern arising from the Co\,2p$_{3/2}$ core level, acquired from the $(3+\delta)+$ peak with photon energy h$\nu$\,=\,1150\,eV, is shown in Fig.~\ref{fig:XPD}(c). Here, again, forward scattering is favoured. From these three data sets, we obtain the out-of-plane parameters of the crystal lattice from the best-fit structural model to the XPD data, keeping the in-plane parameter in the simulation fixed to the result from LEED. We note that LEED gives sharp diffraction spots, consistent with the presence of well-aligned rotational domains of CoO$_2$. We start by assuming that SL CoO$_2$ has a crystalline structure related to the CdI$_2$ type, in accordance with previous predictions \cite{Walton:2015}. The CdI$_2$ structure, however, has only three-fold symmetry. We therefore assume that the nearly six-fold symmetry of the XPD patterns derives from the presence of exactly two domains in the sample, and these are related to each other by mirror symmetry (equivalent to 60$\degree$ in-plane rotation).

The R-factor for the simulations presented in Fig.~\ref{fig:XPD}(a--c) is shown at the bottom right of each panel.  The low values in each case indicate good agreement between the measurement and the simulation. $z_1$ and $z_2$ were determined by optimizing $R$ for all three simulations simultaneously, with the assumption that CoO$_2$ is present on the sample in two equally distributed orientations rotated with respect to one other by 60$\degree$.  The results suggest a slightly smaller distance between Co and top O atoms ($z_1$\,=\,0.95(10)\,{\AA}) than between Co and bottom O atoms ($z_2$\,=\,1.00(10)\,{\AA}).  The uncertainties in $z_1$ and $z_2$ are rough estimates of our confidence in identifying the minimum of each of them in terms of all three parameters simultaneously.  (We note that these uncertainties span a range of $z_1$ and $z_2$ within which the average $R$ value for the three simulations falls mostly within $\Delta R$ of its minimum, although the average $R$ increases more rapidly in the direction of increasing $z_1$ than in the other directions. In any case, the average $R$ might not be the most useful way to think about uncertainty in this particular case.)  Our values of $z_1$ and $z_2$ are consistent with the asymmetric structure proposed by Walton, \foreign{et al.} \cite{Walton:2015}.  In a qualitative sense, such an asymmetry is not unexpected, considering that one side of the layer is at the interface with the substrate, while the other side is hydroxylated and at the interface with vacuum.  
However, the range of our uncertainty does also permit the possibility that $z_1=z_2$, which would not affect any of the conclusions we draw in the present study
\cite{Supplement}.
The atomic structure derived from the XPD simulations is shown from top and side views in Fig.~\ref{fig:XPD}(d), and from an oblique view in Fig.~\ref{fig:XPD}(e). A hexagonal layer of Au atoms corresponding to the unreconstructed (111) surface is shown in Fig.~\ref{fig:XPD}(e) for illustrative purposes, even though the actual XPD simulations do not include the substrate. Because of the lattice mismatch between the SL and the Au(111) substrate, there are many different adsorption sites on the substrate lattice, and thus there is no simple geometric relationship between the emitters in the SL CoO$_2$ and the Au(111); this justifies neglect of the Au(111) in the simulations.

Previously published work has found that adsorbed H at the top of the SL assumes a partially disordered ``labyrinth" structure \cite{Fester:2017b}. In the present study, we observe weak modulation of the hydroxylated O\,1s core-level peak:  this might arise from small structural changes associated with hydroxylation.  However, it is small compared with the modulation of the non-hydroxylated component, and we have not been able to analyse it successfully.\\

\textbf{Electronic structure simulations.}

Here we analyse the electronic structure of the CoO$_2$ SL. To take into account the electron-correlation effects, we utilize the DFT+GA method~\cite{Fang,Ho,Our-PRX}. 
Specifically, we have utilized the implementation of \cite{Our-PRX,Lanata2016},
using the DFT code Wien2k~\cite{WIEN2k}, employing the Local Density Approximation (LDA) and the standard ``fully localized limit" form for the double-counting functional~\cite{LDA+U}.
Our calculations were performed using a $24\times 24$ $k$-point grid and setting the product of the smallest atomic-sphere radius times the largest plane-wave momentum ($RKmax$) to 9.
As in Ref.~\cite{npj-lanata}, we set
the Hund's coupling constant to $J=0.9$~eV, while 
we set the screened Hubbard interaction parameter to $U=6$~eV.

The calculations were performed with the lattice parameters that we measured from LEED and XPD. In the supplementary material~\cite{Supplement} we also show that small changes in the out-of-plane parameters---within the precision of XPD measurments---lead to band structures that are also consistent with the ARPES data. In particular, this includes the inversion-symmetric lattice structure with $z_1=z_2=1.0\,\hbox{\AA}$.

\begin{figure}
\includegraphics[width=0.47\textwidth]{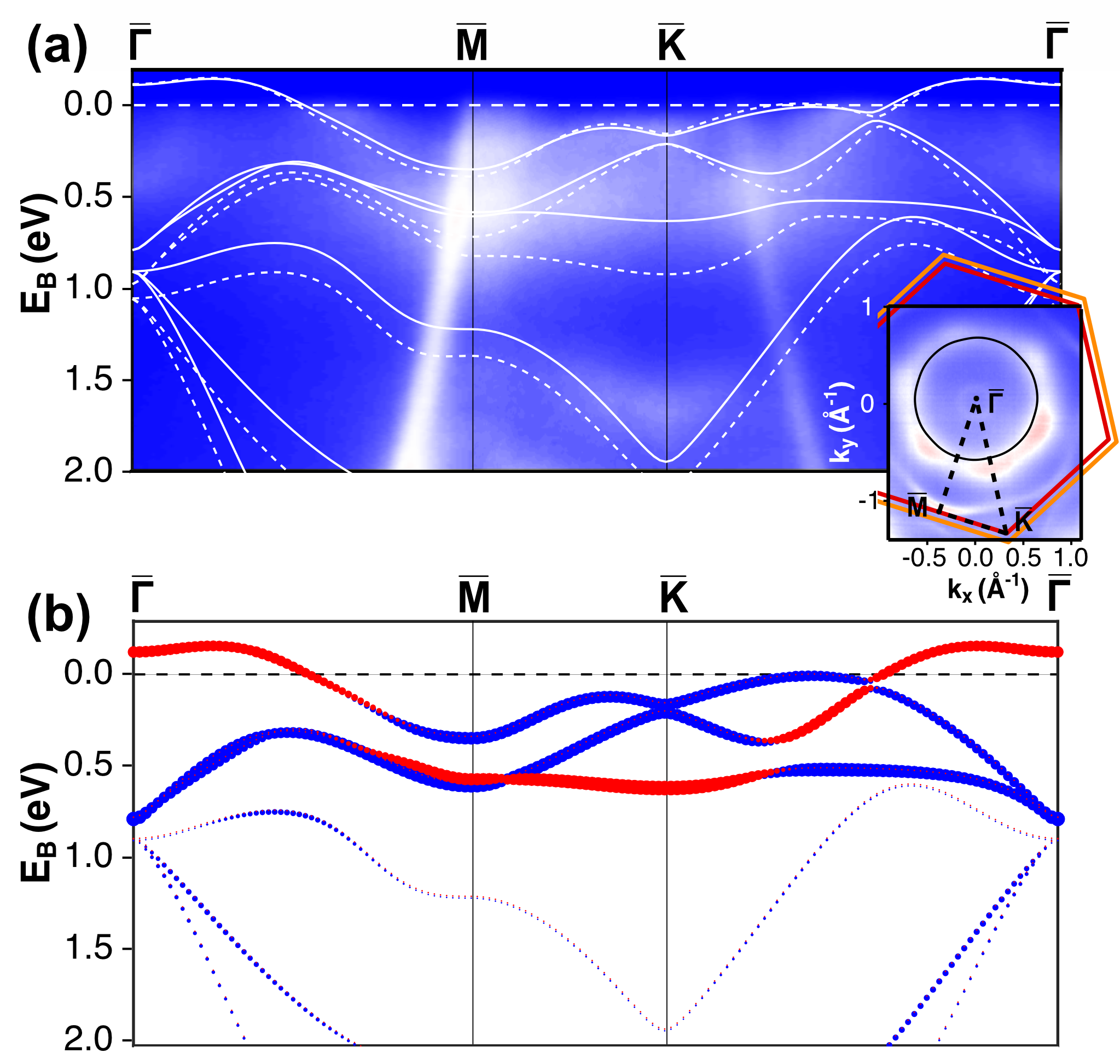}
\caption{(a) Band structure of SL CoO$_2$ calculated with DFT+GA (solid lines) and DFT (dashed lines), superimposed on measured ARPES spectra. The inset shows the corresponding SBZ of Au(111) (CoO$_2$) in orange (red) and the  DFT+GA constant energy contour (black) in comparison with ARPES. 
The theoretical Fermi level is shifted by 90 meV above the calculated value (see discussion in the main text).
(b) Orbital character of the d-electron degrees of freedom within DFT+GA ($Z_{A_1}$ in red, $Z_{1,E}$ and $Z_{2,E}$ in blue).
}
\label{fig:DFT}
\end{figure}

In Fig.~\ref{fig:DFT}(a) we compare the DFT and the DFT+GA bands with the dispersion measured by ARPES.
Note that in Fig.~\ref{fig:DFT} the theoretical Fermi level of the pure single layer is shifted by approximately 90 meV above the calculated value,
as such shift results in a
more satisfactory agreement with the experiments.
In the supplemental material we argue, based on DFT calculations, that this energy shift may be caused by the adsorption of H atoms on the SL~\cite{Supplement}.
We observe that the correlation effects captured by DFT+GA improve the agreement with the experiments considerably compared to bare DFT.

Within the multi-orbital GA framework, the correlation effects on the band structure (Fig.~\ref{fig:DFT}(b)) are encoded in a linear momentum-independent self energy for the Co\,3d electrons, represented as follows~\cite{Our-PRX,Gebhard-FL}:
\begin{eqnarray}
\Sigma(\omega)&=& - \omega\,\frac{I-Z}{Z}
+\Sigma_0
\,.\label{sigma}
\end{eqnarray}
Eq.~\eqref{sigma} is characterized by: (i) the so-called ``matrix of quasi-particle weights" $Z$, whose eigenvalues measure the degree of correlation of the corresponding degrees of freedom, and (ii)
the frequency-independent component of the self-energy matrix $\Sigma_0$, inducing interaction-driven d-electron on-site level shifts.
The Co\,3d manifold is generated by 1 one-dimensional $A_1$ irreducible representation and 2 two-dimensional $E$ representations.
From standard group-theoretical considerations~\cite{Lanata2016} it follows that $Z$ has a non-degenerate $A_1$ eigenvalue $Z_{A_1}$ and 
two two-fold degenerate $E$ eigenvalues $Z_{1,E}, Z_{2,E}$.

Based on our calculation, $Z_{A_1}\simeq 0.68$, 
$Z_{1,E}\simeq 0.72$ and $Z_{2,E}\simeq 0.82$,
indicating that the SL CoO$_2$ system is a moderately-correlated metal.
In Fig.~\ref{fig:DFT}(b) we show the Co\,3d spectral weight of the bands, resolved with respect to the corresponding $p3m1$ point symmetry-group representations.
As expected, the correlation effects are particularly important for capturing the experimental behavior of the d-electron bands, such as those at low binding energy close to the $\overline{\mbox{K}}$-point, which have predominantly $A_1$ character.
However, the d-electron correlations considerably influence the whole band structure, including the O\,2p band  
(because of hybridization effects).

\section{IV. DISCUSSION}

Key questions in 2D-materials research are how the electronic properties of SL systems differ from those of related 3D (bulk) compounds, and how those properties are affected by the environment---for example, by the presence of the substrate, or by adsorbates.  Here we consider these questions for the case of SL CoO$_2$.\\

\textbf{Influence of Au(111) substrate.}

The Au(111) substrate plays a role in catalyzing the growth of CoO$_2$ and in stabilizing the SL \cite{Fester:2017a,Fester:2018b}. Furthermore, as discussed above, the Au(111) substrate introduces an asymmetry in the out-of-plane direction---not only by interacting directly with the SL, but also by inhibiting hydroxylation at the bottom of the layer.  (Full hydroxylation of the top and bottom of the layer---\foreign{i.e.,} synthesis of SL Co(OH)$_2$ on Au(111)---has been shown to be possible in the presence of ``high'' pressures (\foreign{i.e.,} $\sim$10\,mbar) of H$_2$O vapor \cite{Fester:2018b}; however, to our knowledge, hydroxylation at the bottom of the SL on Au(111) does not occur due to the presence of chamber rest gas alone.)

Nevertheless, despite the impact it has on structure, the Au(111) substrate appears to have a weak influence on the electronic properties of the SL.  The most important observation is the fact that calculations neglecting the substrate agree well with the measured ARPES spectra. It is also interesting to observe that the splitting between the two O\,1s XPS peaks, which derive from photoemission from the top and the bottom of the layer, is so small that it cannot be resolved by our measurements.  If the interaction were strong between the substrate and the oxygen atoms at the bottom of the layer, one might expect to see a significant peak shift, similar to what has been observed for MoS$_2$/Au(111) \cite{Bana:2018}, and which we do not see in the present case.\\

\textbf{Hydroxylation.}

To estimate the amount of hydrogen present, we note that in the O\,1s core level of Fig.~\ref{fig:XPS}(a) the hydroxyl component accounts for 43\% of the combined intensity of the two peaks. We assume that there is attenuation of the photoemission intensity from the bottom O layer (due to an inelastic mean free path of $\lambda$\,=\,5.3\,{\AA} \cite{Quases, Shinotsuka:2015}) but that the presence of H does not attenuate the photoemission intensity. Furthermore, we assume that hydroxylation happens only at the top of the CoO$_2$ layer \cite{Fester:2017b}, and that each H is bonded only to a single O. This results in a value of 57\% hydroxylated O at the top of the layer.  Note, however, that there is a rather high level of uncertainty in this estimate, depending on the validity of our several assumptions. Also, this rough estimate does not take into account the persistance of local regions of CoO after the second oxidation step: these certainly persist on the sample surface, as can be seen in Fig.~\ref{fig:XPS}.  We have observed---in agreement with previous literature \cite{Fester:2018a,Fester:2018b}---that CoO hydroxylates less extensively than CoO$_2$; thus, our estimate represents only a lower bound for the extent of hydroxylation in our sample.  Our findings appear roughly consistent with those of previous studies, which have shown that hydroxylation of the CoO$_2$ layer is approximately $\frac{2}{3}$ (\foreign{i.e.,} two out of every three O atoms at the top of the layer hydroxylated) after storage in UHV conditions for several hours \cite{Fester:2017b}.  Indeed, the sample that generated the XPS data shown above was stored under UHV conditions for a few days before the measurements in Fig.~\ref{fig:XPS} were acquired.

By contrast, the samples that generated the ARPES data above were stored in UHV conditions for only a few hours before they were measured.  They might, therefore, be less hydroxylated than the samples measured with XPS. However, on the basis of the existing literature we would not expect the level of hydroxylation to be less than approximately $\frac{1}{3}$, as this is the lower bound previously observed in freshly-grown samples \cite{Fester:2017b}. The measured Fermi contour in Fig.~\ref{fig:ARPES}  is 70(3)\,\% filled, which indicates a charge transfer to the SL of 0.40(6)e per unit cell (by contrast with undoped CoO$_2$, whose Fermi contour would be half-filled). Thus, the amount of hydroxylation that we estimate using XPS would likely be sufficient to generate the charge transfer that we observe in ARPES, even without charge transfer from the Au(111) substrate.

An obvious question is what influence the hydroxylation has on the electronic dispersion. Previous work \cite{Fester:2017b} has shown that the H ions are not well ordered on the surface;  therefore, we might expect the OH groups to shift and broaden the band structure rather than to impact the bare-band dispersion. Indeed, the calculations shown in Fig.~\ref{fig:DFT} successfully reproduce most of the main features of the band structure without the inclusion of hydroxylation. Nevertheless, there could, of course, be some level of weak ordering of the H ions, and we consider the implications of this for the band dispersion in \cite{Supplement,kresse1996efficiency,kresse1999ultrasoft,perdew1996generalized,monkhorst1976special,Demaison:2007}.\\

\textbf{Comparison of SL CoO$_2$ with related bulk materials.}

In light of the apparently weak impact of the substrate on the electronic dispersion, and the likelihood of little if any charge transfer from the substrate, it seems reasonable to consider our CoO$_2$ system as approximately ``2D.''  Here, then, we discuss the electronic structure of this 2D material in relation to analogous layered bulk systems, such as Na$_x$CoO$_2$ and Na$_x$CoO$_2\cdot$\,yH$_2$O.
This is a particularly interesting point of consideration, because the
superconducting and magnetic properties of the layered bulk materials have been interpreted as arising from the``pseudo-2D'' nature of the weakly interacting CoO$_2$ atomic layers: the layers are partially isolated from one another in bulk materials by interleaving layers of Na$^+$, H$_2$O, and H$_3$O$^+$ \cite{Takada:2003,Sakurai:2015,Lorenz:2003, Milne:2004, Wang:2005}. (We note that superconductivity would not be expected in our samples, because of the proximity effect of the Au(111) substrate.)

The electronic dispersion of the SL is remarkably similar to that of related layered bulk materials \cite{Singh:2000, Lee:2004, Hasan:2004, Arakane:2008, Yang:2005, Qian:2006}. The large hole pocket around $\overline{\Gamma}$ is the most obvious feature of the bulk dispersion, and this is true for the SL, as well. Discussion in the literature has surrounded the question of why six small hole pockets that are predicted to cross the Fermi level along $\overline{\Gamma}\overline{\mbox{K}}$ in the bulk dispersion \cite{Singh:2000, Lee:2004} are not observed experimentally at any doping level \cite{Hasan:2004,Yang:2004,Yang:2005,Arakane:2008}.  Our results are similar to those of ARPES studies of the bulk, in that no hole pockets rise above the Fermi level along $\overline{\Gamma}\overline{\mbox{K}}$.  However, we do note that there are hole pockets just below the Fermi level, and that the measured Fermi contour of the SL exhibits diffuse, elongated intensity in the $\overline{\Gamma}\overline{\mbox{K}}$ direction, due to the presence of these shallow states located in close vicinity to, and leaking some intensity across, the Fermi level.

Comparing the properties of the bulk and the SL, it is worth pointing out that there are some structural differences between our samples and the stacked layers of similar bulk systems.  The in-plane lattice constant we have identified here, 3.12(3)\,{\AA}, is larger than that reported for the bulk (2.8222(13)\,{\AA}~\cite{Amatucci:1996}). The O height above Co as reported for the bulk (\textit{e.g.,} 0.91\,{\AA} for NaCo$_2$O$_4$~\cite{Singh:2000}) is within the range of the uncertainty of our measurements of the SL, but the bulk layer would be expected to be symmetrical around the Co plane, whereas our results agree with earlier suggestions that the SL is likely to be asymmetrical around the Co plane~\cite{Walton:2015}.\\

\textbf{Kinks and electron-phonon coupling.}

A final point of interest relates to electron-phonon coupling, a topic relevant to superconductivity. Several studies of related bulk CoO$_2$-based materials identify kinks in the bands that cross the Fermi level \cite{Hasan:2004, Geck:2007, Arakane:2008}. In the case of the SL, although at first glance the data in Fig. \ref{fig:ARPES}(a) does appear to exhibit kinks at both band crossings, we do not find any decisive indication that such kinks definitely occur.  As can be seen from Fig.~\ref{fig:DFT}, the sharp downturn in the filled band dispersion just below the Fermi level along the $\overline{\Gamma}\overline{\mbox{K}}$ direction seems adequate to explain the shape of the band there, in agreement with Qian \textit{et al.} \cite{Qian:2006}. Along the $\overline{\Gamma}\overline{\mbox{M}}$ we were not able to convincingly fit any kink in the  dispersion, so the situation here remains somewhat unclear. 


\section{V. CONCLUSIONS}

Using XPD, we have determined the crystalline structure of SL CoO$_2$ on Au(111), finding good agreement with previous predictions. We have reported the electronic structure on the basis of ARPES measurements and DFT+GA calculations. Our calculations describe the ARPES data well, and indicate that SL CoO$_2$ is characterized by moderate electronic correlations.  We have observed significant hydroxylation at the top of the layer, and found that the main effect of the H impurities is a shift of the Fermi level.  Our results suggest a weak interaction between the SL and the Au(111) substrate.

\section{ACKNOWLEDGMENTS}
We gratefully acknowledge financial support from the VILLUM FONDEN via the Centre of Excellence for Dirac Materials (Grant No. 11744) and VILLUM project grant no. 13264 (J.V.L.). Y.-X.Y. was supported by the U.S. Department of Energy, Ofﬁce of Science, Basic Energy Sciences, as part of the Computational Materials Science Program.  C.E.S. received support from the European Community's Seventh Framework Programme (FP7/2007-2013) CALIPSO under grant agreement no 312284.

\theendnotes

\bibliographystyle{apsrev}
\bibliography{CoO2.bib}

\end{document}


\title{\textit{Supplementary Material}\\
Electronic Properties of Single-Layer CoO$_2$/Au(111)}

\author{Ann Julie U. Holt}
\affiliation{Department of Physics and Astronomy, Interdisciplinary Nanoscience Center (iNANO), Aarhus University, 8000 Aarhus C, Denmark}
\author{Sahar Pakdel}
\affiliation{Department of Physics and Astronomy, Interdisciplinary Nanoscience Center (iNANO), Aarhus University, 8000 Aarhus C, Denmark}
\author{Jonathan R. Fernandez}
\affiliation{Department of Chemistry, Interdisciplinary Nanoscience Center (iNANO), Aarhus University, 8000 Aarhus C, Denmark}
\author{Yu Zhang}
\affiliation{Central Laser Facility, STFC Rutherford Appleton Laboratory, Didcot, Oxfordshire OX11 0QX, UK}
\author{Davide Curcio}
\affiliation{Department of Physics and Astronomy, Interdisciplinary Nanoscience Center (iNANO), Aarhus University, 8000 Aarhus C, Denmark}
\author{Paolo Lacovig}
\affiliation{Elettra - Sincrotrone Trieste S.C.p.A., Trieste, Italy}
\author{Yong-Xin Yao}
\affiliation{Ames Laboratory-U.S. DOE and Department of Physics and Astronomy, Iowa State University, Ames, Iowa 50011, USA}
\author{Jeppe V. Lauritsen}
\affiliation{Department of Chemistry, Interdisciplinary Nanoscience Center (iNANO), Aarhus University, 8000 Aarhus C, Denmark}
\author{Silvano Lizzit}
\affiliation{Elettra - Sincrotrone Trieste S.C.p.A., Trieste, Italy}
\author{Nicola Lanata}
\affiliation{Department of Physics and Astronomy, Interdisciplinary Nanoscience Center (iNANO), Aarhus University, 8000 Aarhus C, Denmark}
\author{Philip Hofmann}
\affiliation{Department of Physics and Astronomy, Interdisciplinary Nanoscience Center (iNANO), Aarhus University, 8000 Aarhus C, Denmark}
\author{Marco Bianchi}
\affiliation{Department of Physics and Astronomy, Interdisciplinary Nanoscience Center (iNANO), Aarhus University, 8000 Aarhus C, Denmark}
\author{Charlotte E. Sanders}
\email{charlotte.sanders@stfc.ac.uk}
\affiliation{Central Laser Facility, STFC Rutherford Appleton Laboratory, Didcot, Oxfordshire OX11 0QX, UK}

\date{\today}
\maketitle

\fontsize{12}{18}\selectfont

\noindent \textbf{1. Surface coverage determinations from scanning tunneling microscopy (STM) and x-ray photoelectron spectroscopy (XPS)}\\

\begin{figure}
\includegraphics[width=\textwidth]{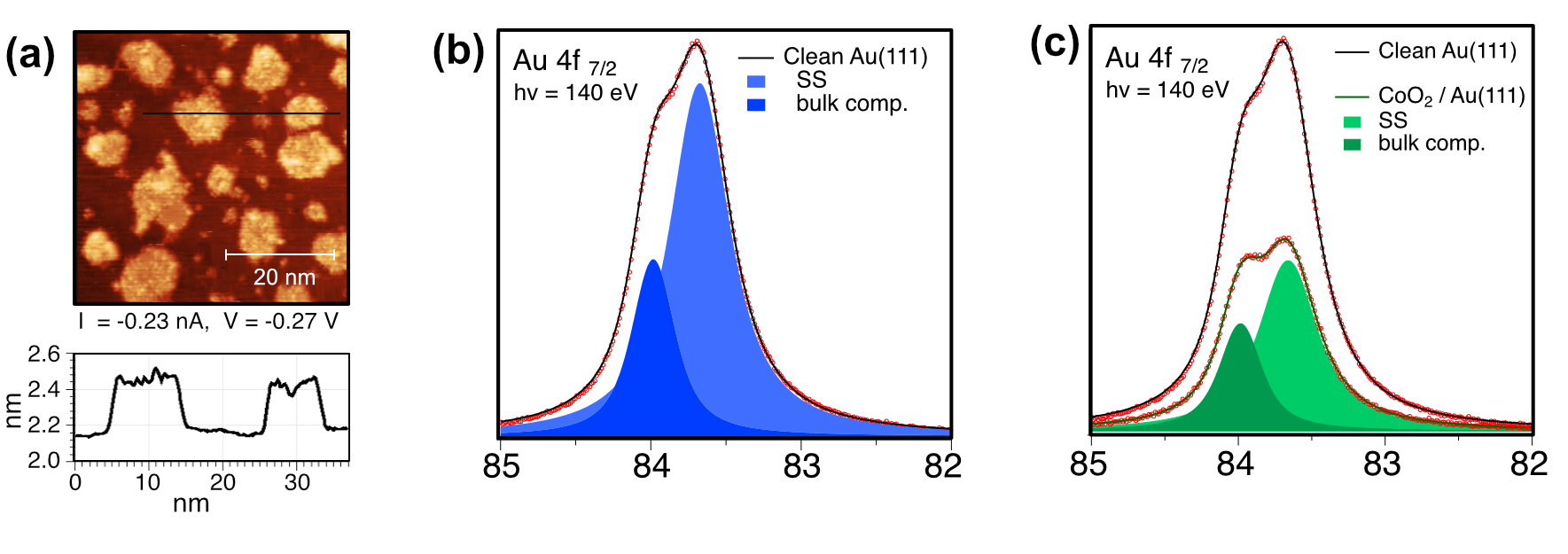}
\caption{STM and XPS measurements assessing sample coverage. (a) STM data, from the SGM3 beamline, showing fully oxidised SL CoO$_2$ islands.  (b,c) Au\,4f$_{7/2}$ core-level spectrum of the clean substrate (b) and after CoO$_2$ growth (c). We are able to fit the spectrum with exactly two peaks; any possible third component that might arise from interaction between the Au(111) surface and the oxide SL is not resolvable in the data. These XPS measurements were made at the SuperESCA beamline, h$\nu$ = 140\,eV, at normal emission. See main text for additional measurement parameters.} 
 \label{fig:S1}
\end{figure}

\noindent The surface coverage of samples fabricated for angle-resolved photoemission spectroscopy (ARPES) characterisation was estimated using STM, as shown in Figure \ref{fig:S1}\,(a). The sample in the figure was fabricated according to the procedure described in the main text, with 30-min. Co deposition and subsequent CoO-to-CoO$_2$ oxidation for 2\,hrs., resulting in a surface coverage of approximately 50\%. The measurement shows islands of 10--15\,nm size, with 3.0\,$\pm$\,0.2\,{\AA} apparent height at the scanning parameters shown, consistent with earlier reports \cite{Walton:2015, Fester:2017a}. It should be noted that the STM apparent height reflects both topographic and electronic differences between the adlayer and the substrate, and was not used to determine layer thickness or spacing from the Au substrate. 

The surface coverage of the sample used for XPD measurements was assessed \foreign{in-situ} through analysis of high-resolution Au\,4f$_{7/2}$ core-level spectra.  Examples are shown in Fig. \ref{fig:S1}\,(b) and (c). The clean Au(111) substrate is presented in Fig. \ref{fig:S1}\,(b), with the Au bulk and surface-shifted (SS) \cite{Heimann:1981} peak components shown in dark and light blue, respectively. The majority of the measured intensity initially arises from the SS component. After the CoO$_2$ growth procedure, the SS component is attenuated by 55\,\%, as shown in Fig. \ref{fig:S1}\,(c). A rough estimate of the surface coverage was made by assuming that CoO$_2$ quenches the surface binding energy shift within the local region where the substrate is covered.  Extrapolating from this, we estimate the surface coverage to be 55\%. We note also that the bulk component is attenuated by 40\,\% once the second oxidation step in the fabrication process (CoO-to-CoO$_2$ transition) is completed, and that the SS component remains approximately constant during this step. This pair of observations is consistent with the conclusion that the CoO$_2$ structure attenuates photoemission intensity from the covered areas of the substrate (thus suppressing the bulk component) without affecting the intensity arising from the bare areas (which are the source of the SS peak intensity). We assume on the basis of our STM measurements that the extent of the coverage of the CoO islands is the same as that of the CoO$_2$ islands.  \\

\noindent\textbf{2. Fitting XPD measurements}\\

The reliability factor, defined in the main text, was used to optimise the agreement between the experimental and the simulated XPD patterns. Diffraction patterns were calculated for a variety of structural configurations and parameters, with a structure similar to the distorted octahedral crystal structure assumed as an initial guess, on the basis of earlier-published work \cite{Walton:2015}. On account of the six-fold symmetry of the XPD patterns, the model assumes an incoherent superposition of two mirror domains. The total photoemission intensity is thus described as

\begin{equation}
    I = AI_1 + (1-A)I_2,
\end{equation}

where I$_1$ and I$_2$ are the intensity contributions from the two distinct mirror domains (domain 1 and domain 2, respectively) and A is a weight factor quantifying the presence of each domain. The best-fit simulated XPD pattern was found by running an R-factor optimisation in a five-dimensional parameter space, independently changing the weight factor, the sample Debye temperature, the muffin-tin inner potential, and the crystal structure parameters z$_1$ and z$_2$ (parameters z$_1$ and z$_2$ are defined in the main text). The optimised values for the weight factor, the sample Debye temperature, the muffin-tin inner potential, was 0.5, 300\,K and 15\,eV, respectively. The in-plane lattice parameter was determined by LEED measurements and therefore not included in the R-factor optimisation parameter space. \\

\noindent\textbf{3. Hydroxylation effect}\\

In Ref.~\cite{Fester:2017b} it is suggested that, shortly after growth, a $\sqrt{3}\times\sqrt{3}$ superstructure is formed on the CoO$_2$ sample surface, due to adsorption of H atoms.
Here we investigate the influence of H adsorption on the band structure of CoO$_2$. For simplicity, in our calculations we consider ordered structures.
Specifically, to infer the effect of adsorbate coverage similar to that which was observed experimentally in the present study, we consider a commensurate configuration with coverage $\frac{1}{3}$ (as suggested in Ref.~\cite{Fester:2017b}), as well as a commensurate configuration with coverage $\frac{2}{3}$.

\begin{figure}
\includegraphics[width=0.9\textwidth]{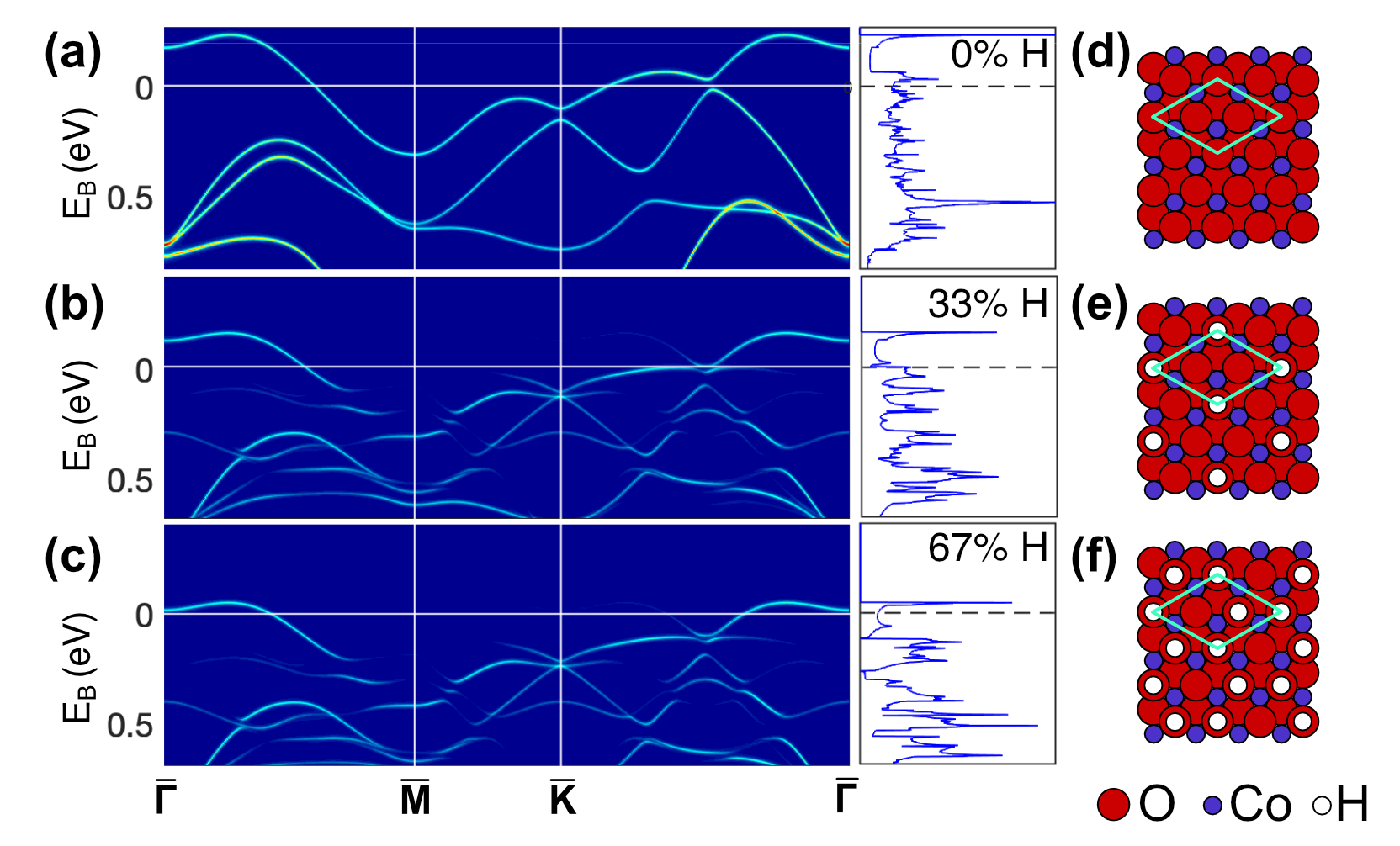}
\caption{(a-c) DFT band structure and DOS of $\sqrt{3}\times\sqrt{3}$ CoO$_2$ superstructure with 0, $\frac{1}{3}$ and $\frac{2}{3}$ H coverage, respectively. The bands have been unfolded on the high symmetry path of the CoO$_2$ primitive BZ. (d-f) The schematic of the supercells used in (a-c) and defined based on Ref.~\cite{Fester:2017b}.} 
 \label{fig:S2}
\end{figure}

In the following band-structure calculations we set the O-H distance to $h=0.98$\,{\AA}---which is the distance that we calculated by minimizing the total energy for the $\frac{1}{3}$-coverage case.
This value is in good agreement with the literature \cite{Demaison:2007}.
Performing a supercell calculation and unfolding the bands on the high-symmetry path of the original BZ of CoO$_2$, we observe that the overall behaviour of the bands is conserved.
As shown in Fig.~\ref{fig:S2}, the main effects of the H impurities are a Fermi-level shift and the occurrence of mini-bands. 
We observe that, for $\frac{2}{3}$ coverage, the Fermi energy of the CoO$_2$ layer is effectively shifted by about 100 meV, resulting in a more satisfactory agreement between theory and experiments. 
Note that the effect of H is introduced in the DFT bands of Fig.~4 via a Fermi level shift. On the other hand, the coverage of the randomly distributed H atoms in the actual samples of Fig.~4 are experimentally estimated to be $\sim 57\,\%$ (less than $2/3$). Hence, a slightly smaller Fermi energy shift ($\sim 90\,$meV) is expected.\\


The calculations of this specific section were performed on supercells with bare DFT (no correlation effects included). Since the plane-wave basis is more convenient for unfolding the band structure, we utilized Vienna Ab initio Simulation package (VASP) (while Wien2k code was used in the main text).
%
The exchange–correlation potentials were described through the Perdew–Burke–Ernzerhof (PBE) functional within the generalised gradient approximation (GGA) formalism~\cite{perdew1996generalized}. A plane wave basis set was used with a cutoff energy of 400\,eV on a $24\times24$ Monkhorst-Pack \cite{monkhorst1976special} $k$-point mesh. A vacuum region of 18\,{\AA} along the $z$-direction (orthogonal to the layer plane) was used to separate the layers in order to minimise the interaction between the periodic repetitions of the cell. The spin-orbit coupling was not included in these calculations.\\

\noindent\textbf{4. Effect of varying lattice parameters}\\

\begin{figure}
\includegraphics[width=0.7\textwidth]{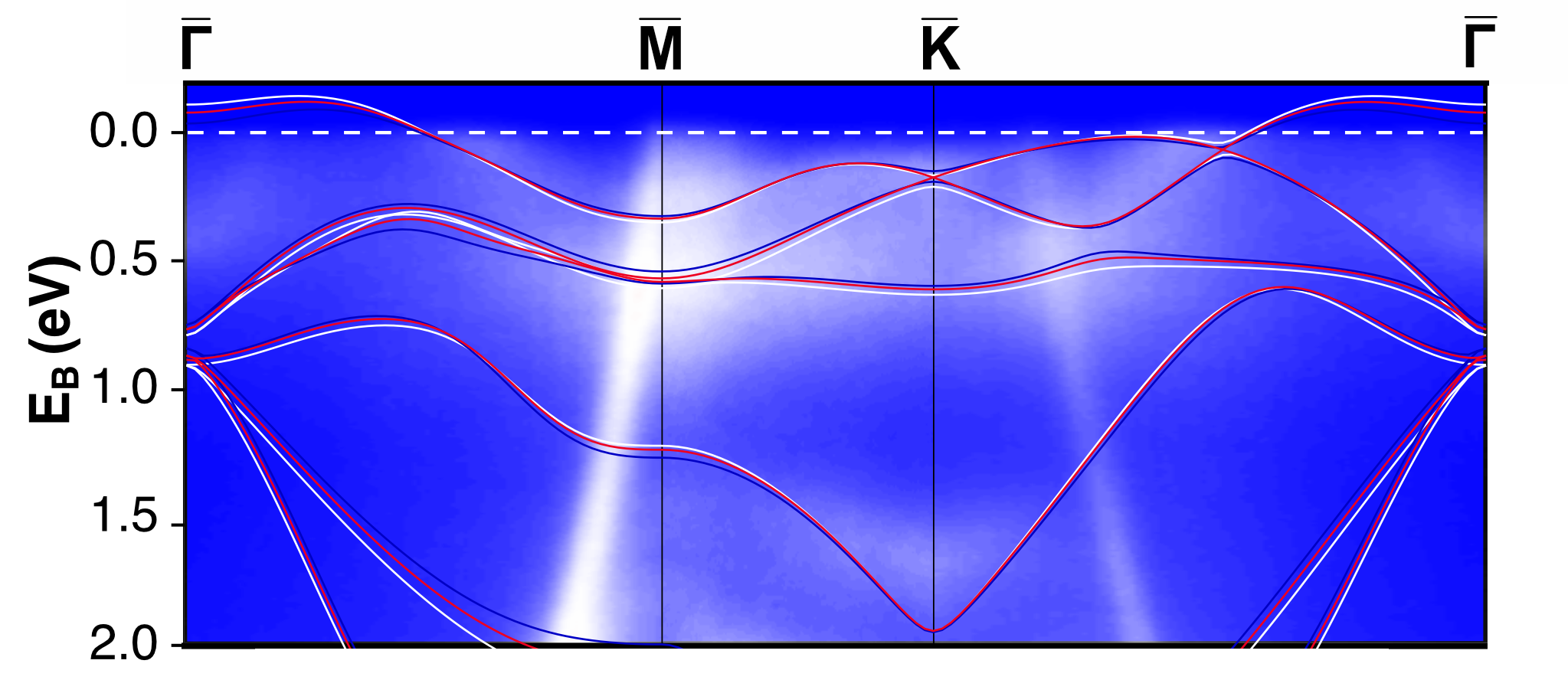}
\caption{Calculated band structures of CoO$_2$, including correlation effects, overlaid on the ARPES data from Fig.\,1 of the main text.  Red: $z_1=z_2=1.0$\,{\AA}.  Blue:  $z_1=1.0$\,{\AA}, $z_2=1.05$\,{\AA}.  White: $z_1=0.95$\,{\AA}, $z_2=1.0$\,{\AA}. In-plane lattice constant is 3.12\,{\AA} for all three calculations.} 
 \label{fig:S3}
\end{figure}

Since the precision of the XPD measurements for the out-of-plane lattice parameters is less than the precision of LEED for the in-plane lattice constant, in Fig.~\ref{fig:S3} we study the sensitivity of the calculated band structure to variations in out-of-plane lattice parameters ($z_1, z_2$) within the limits of precision of the XPD measurement. Small changes can be observed in a comparison between the $z_1=z_2=1.0$\,{\AA} inversion-symmetric case (red); the $z_1=1.0$\,{\AA}, $z_2=1.05$\,{\AA} case (blue); and the XPD best-fit case $z_1=0.95$\,{\AA}, $z_2=1.0$\,{\AA} (white). However, all of them are compatible with the ARPES bands.

\bibliographystyle{apsrev}
\bibliography{CoO2.bib}